\documentclass[useAMS,usenatbib,usegraphicx,referee]{mn2e}
\usepackage{txfonts}

\newcommand{\ko}{kr_{L,0}}

\title[A kinetic approach to growth rates of streaming instability at supernova
shocks]
{A kinetic approach to cosmic ray induced streaming instability at supernova
shocks}
\author[E. Amato and P. Blasi]{E. Amato$^{1}$\thanks{E-mail:
amato@arcetri.astro.it} 
and P. Blasi$^{1,2,3}$\thanks{E-mail: blasi@arcetri.astro.it}\\
$^{1}$INAF-Osservatorio Astrofisico di Arcetri, 
Largo E. Fermi, 5, 50125, Firenze, Italy\\
$^{2}$Fermilab/Center for Particle Astrophysics, USA\\
$^{3}$INFN/Laboratori Nazionali del Gran Sasso, S.S. 17 BIS km. 18.910
67010 Assergi, L'Aquila, Italy}
\begin{document}

\date{Accepted ----. Received -----}


\maketitle

\label{firstpage}

\begin{abstract}
We show that a purely kinetic approach to the excitation of waves
by cosmic rays in the vicinity of a shock front leads to predict the
appearance of a non-alfv\'enic fastly growing mode which has the same
dispersion relation as that previously found by \cite{bell04} by treating the
plasma in the MHD approximation. The kinetic approach allows us to
investigate the dependence of the dispersion relation of these waves on the 
microphysics of the current which compensates the cosmic ray flow.
We also show that a resonant and a non-resonant mode may appear at the same
time and one of the two may become dominant on the other depending on the
conditions in the acceleration region. We discuss the role of the unstable
modes for magnetic field amplification and particle acceleration in supernova
remnants at different stages of the remnant evolution.
\end{abstract}

\begin{keywords}
acceleration of particles - shock waves
\end{keywords}

\section{Introduction}
\label{sec:intro}

The problem of magnetic field amplification at shocks is central to
the investigation of cosmic ray acceleration in supernova remnants. The
level of scattering provided by the interstellar medium turbulent
magnetic field is insufficient to account for cosmic rays with energy
above a few GeV, so that magnetic field amplification and large scattering
rates are required if energies around the knee are to be reached. The
chief mechanism which may be responsible for such fields is the
excitation of streaming instabilities (SI) by the same particles which are
being accelerated (\cite{skilling75c,bell78a,lc83a,lc83b}). The effect of
magnetic field amplification on the maximum energy reachable at supernova
remnant (SNR) shocks was investigated by \cite{lc83a,lc83b}, who
reached the conclusion that cosmic rays could be accelerated up to 
energies of order $\sim 10^4-10^5$ GeV at the beginning of the Sedov phase. 
This conclusion was primarily based on the assumption of Bohm diffusion 
and a saturation level for the induced turbulent field $\delta B/B\sim 1$. 
On the other hand, recent observations of the X-ray surface brightness of the 
rims of SNRs have shown that $\delta B/B\sim 100-1000$ (see \cite{volk} for
a review of results), thereby renewing the interest in the
mechanism of magnetic field amplification and in establishing its
saturation level.  
It is however worth to recall that the interpretation of the
X-ray observations is not yet unique: the narrow
rims observed in the X-ray synchrotron emission could be due to the
damping of the downstream magnetic field \cite{pohl} rather than to
severe synchrotron losses of very high energy electrons, although
this interpretation has some serious shortcomings (see \cite{morlino}
for a discussion).

In this context of excitement, due
to the implications of these discoveries for the origin of cosmic rays,
\cite{bell04} discussed the excitation of modes in a plasma treated
in the MHD approximation and found that a new, purely growing,
non-alfv\'enic mode appears for high acceleration efficiencies. The author
predicted a saturation of this SI at the level $\delta B/B\sim
M_A (\eta v_s/c)^{1/2}$ where $\eta$ is the cosmic ray pressure in units
of the kinetic pressure $\rho v_s^2$, $v_s$ is the shock speed and
$M_A=v_s/v_A$ is the Alfv\'enic Mach number. For comparison, 
standard SI for resonant wave-particle interactions leads to expect 
$\delta B/B\sim M_A^{1/2} \eta^{1/2}$. For efficient acceleration
$\eta\sim 1$, and typically, for shocks in the interstellar medium,
$M_A\sim 10^4$. Therefore Bell's mode leads to $\delta B/B\sim
300-1000$ while the standard SI gives $\delta B/B\sim 30$. It is also
useful to notice that the saturation level predicted by \cite{bell04}
is basically independent of the value of the background field, since
$\delta B^2/8\pi\sim (1/2)(v_s/c)P_{CR}$, where $P_{CR}$ is the cosmic
ray pressure at the shock surface.  

The resonant and non-resonant mode have different properties in other
respects as well. A key feature consists in the different wavelengths that are
excited. The resonant mode with the maximum growth rate has wavenumber 
$k$ such that $\ko=1$, where $r_{L,0}$ is the Larmor radius of the
particles that dominate the cosmic ray number density at the
shock, namely, for typical spectra of astrophysical interest,
the lowest energy cosmic rays at the shock. At the shock location, the
minimum momentum is the injection momentum, while at larger distances
from the shock, the minimum momentum is determined by the diffusion
properties upstream and is higher than the injection momentum, since
higher energy particles diffuse farther upstream. When the non-resonant
mode exists, its  
maximum growth is found at $\ko\gg 1$. There may potentially be many
implications of this difference: the particle-wave interactions which are
responsible for magnetic field amplification also result in particle
scattering (diffusion). The diffusion properties for resonant and
non-resonant interactions are in general different. The case of
resonant interactions has been studied in the literature
(e.g. \cite{lc83a}), at least for the situation $\delta B/B\ll 1$, but
the diffusion coefficient for  non-resonant interactions (in either
the linear or non-linear case) has not  been calculated (see however
\cite{zira08b}).  
The difference in wavelengths between the two modes, in 
addition to different scattering properties, also suggests that the
damping will occur through different mechanisms.   

The calculation of \cite{bell04} has however raised some concerns
due to the following three aspects: 1) the background plasma was treated in the
MHD approximation; 2) a specific choice was made for the current established in
the upstream plasma to compensate for the cosmic ray (positive) current;
3) the calculation was carried out in a reference frame at rest with the
upstream plasma, where stationarity is in general not realized (although for
small scale perturbations, the approximation of stationarity may be sometimes
justified).  

In the present paper we derive the dispersion relation of the waves in
a purely kinetic approach and investigate different scenarios for the
microphysics that determines the compensating current. We show that
the fastly growing non resonant mode appears when particle
acceleration is very efficient, but whether it dominates over
the well known resonant interaction between particles and alfv\'en modes 
depends on the parameters that characterize the shock front, its Mach
number primarily.

\cite{bell04} also investigated the developement of the non-resonant modes by
using numerical MHD simulations. His results have been recently confirmed by
\cite{zira08a} with a similar approach. \cite{niem08} made a first attempt to
investigate the development of the non-resonant modes by using PIC simulations.
In this latter case, the authors find that the non-resonant mode saturates at
a much lower level than found by \cite{bell04}. However, as briefly discussed 
in \S~\ref{sec:concl}, these simulations use a set up that makes them difficult
to compare directly with Bell's results.

The paper is organized as follows: in \S~\ref{sec:calc} we derive the 
dispersion relation of the unstable modes within a kinetic approach and 
adopting two different scenarios for the compensation of the cosmic ray 
current, namely compensation due to the motion of cold electrons alone 
(\S~\ref{sec:calc1}), and to the relative drift of protons and electrons
(\S~\ref{sec:calc2}); in \S~\ref{sec:modes} we discuss the relative importance
of the resonant and non-resonant modes depending on the physical parameters
of the system; we also derive
analytic approximations for the large (\S~\ref{sec:largek}) and 
small (\S~\ref{sec:smallk}) wavenumber limits; finally in \S~\ref{sec:res}
we study the different modes during the Sedov evolution of a ``typical'' 
supernova remnant and for different assumptions on the background magnetic 
field strength; we conclude in \S~\ref{sec:concl}.

Throughout the paper we use the expressions {\it accelerated particles} and
{\it cosmic rays} as referring to the same concept. 

\section{The kinetic calculation}
\label{sec:calc}

In this section we describe our kinetic calculation of the linear
growth of waves excited by streaming cosmic rays upstream of a
shock. This type of analysis is not suited for the computation of the
saturation level of the instability, but it can be used to investigate
the type of modes that may possibly grow to non linear levels.

In the following, all calculations refer to the shock location and 
not to an arbitrary location in the plasma upstream of the shock. This
is to say that the minimum momentum of the particles considered in our
calculations is the injection momentum.

In the reference frame of the upstream plasma the {\it gas} of cosmic rays
moving with the shock appears as an ensemble of particles streaming 
at super-Alfv\'enic speed. This situation is expected to lead to streaming
instability, as was indeed demonstrated in several previous papers (see
\cite{krall} for a technical discussion). 

In the reference frame of the shock, cosmic rays are approximately
stationary and roughly isotropic. The upstream background plasma moves
with a velocity $v_s$ towards the shock and is made of protons and
electrons. The charge of cosmic rays, assumed to be all protons
(positive charges) is compensated by processes which depend on the
microphysics and need to be investigated accurately.

The x-axis, perpendicular to the shock surface has been chosen to go
from upstream infinity ($x=-\infty$) to downstream infinity
($x=+\infty$). Therefore a cosine of the pitch angle $\mu=+1$
corresponds to particles moving from upstream towards the shock. 

The dispersion relation of waves in this composite plasma, in the
test-particle regime that we wish to investigate here, can be written as
(\cite{krall}):
\begin{equation}
\frac{c^2 k^2}{\omega^2}=1+\sum_\alpha \frac{4\pi^2 q_\alpha^2}{\omega}
\int_0^{\infty} dp \int_{-1}^{+1} d\mu 
\frac{p^2 v(p) (1-\mu^2)}{\omega+k v(p) \mu \pm \Omega_\alpha}\left[
\frac{\partial f_\alpha}{\partial p}+\left( \frac{k
  v}{\omega}+\mu\right)\frac{1}{p} \frac{\partial f_\alpha}{\partial \mu}
\right],
\label{eq:disp}
\end{equation} 
where the index $\alpha$ runs over the particle species in the plasma,
$\omega$ is the wave frequency corresponding to the wavenumber $k$ and
$\Omega_\alpha$ is the relativistic gyrofrequency of the particles of
type $\alpha$, which in terms of the particle cyclotron frequency 
$\Omega_\alpha^*$ and Lorentz factor $\gamma$ is 
$\Omega_\alpha=\Omega_\alpha^*/\gamma$. For the background plasma and for 
any population of cold electrons one has $\Omega_\alpha\approx\Omega_\alpha^*$. 

The positive electric charge of the accelerated cosmic rays, assumed here to be
all protons, with total number density $N_{CR}$,
must be compensated by a suitable number of electrons in the upstream
plasma. In the following subsections we discuss two different ways of
compensating the cosmic ray current and charge. In the first calculation
we assume that there is a population of cold electrons which is at rest 
in the shock frame and drifts together with cosmic rays. These electrons
cancel exactly the positive charge of cosmic rays. This approach is
similar to that of \cite{zweib79,zweib03} and resembles more closely
the assumptions of the MHD approach of \cite{bell04}. In the second 
calculation we assume that the current of cosmic ray protons is
compensated by background electrons and protons flowing at different
speeds. This approach is similar to that of \cite{achter83}. 

\subsection{Model A: cold electrons}
\label{sec:calc1}

Let $n_i$ and $n_e$ be the number density of ions (protons) and
electrons in the background plasma upstream of the shock. In this
section we consider the case in which a population of cold electrons
with density $n_{cold}$ streams together with cosmic rays and compensates 
their charge. Therefore
$n_e=n_i$ and $n_{cold}=N_{CR}$. In terms of distribution functions,
the four components can be described as follows:
\begin{eqnarray}
f_{i}(p,\mu)&=&\frac{n_i}{2 \pi p^2}\delta (p-m_i v_s) \delta (\mu-1) \\
f_{e}(p,\mu)&=&\frac{n_e}{2 \pi p^2}\delta (p-m_e v_s) \delta (\mu-1) \\
f_{e}^{cold}(p)&=&\frac{N_{CR}}{4 \pi p^2} \delta (p)\\
f_{CR}(p)&=&\frac{N_{CR}}{4 \pi} g(p).
\end{eqnarray}
In the latter equation, which describes cosmic rays, $g(p)$ is a function
normalized so that $\int_{p_0}^{p_{max}} dp~p^2 g(p)=1$. In the expressions
above the background ions and electrons have been assumed to be cold (zero
temperature). Introducing the thermal distribution of these particles does not
add, as a first approximation, any important information to the analysis of the 
stability of the modes. One should check however that damping does not
play an appreciable role, especially for the modes with high $k$ (see the paper
by Everett et al. (in preparation)).

The contribution of the background plasma of electrons and protons to the right
hand side of Eq.~\ref{eq:disp} is easily calculated to be:
\begin{equation}
-\frac{4\pi e^2 n_i}{\omega^2 m_i}\frac{\omega+k v_s}{\omega+k v_s\pm
\Omega_i^*} - \frac{4\pi e^2 n_e}{\omega^2 m_e}\frac{\omega+k v_s}{\omega+k
v_s\pm \Omega_e^*}.
\end{equation}
Similarly the cold electrons with density $N_{CR}$ contribute a term:
\begin{equation}
-\frac{4\pi e^2}{\omega}\frac{N_{CR}}{m_e (\omega\pm\Omega_e^*)}.
\end{equation}
The calculation of the cosmic ray contribution is slightly more complex. In
its most general form, it can be written as 
\begin{equation}
\chi_{CR}=\frac{\pi e^2 N_{CR}}{\omega} \int_{0}^{\infty} dp v(p) p^2 \frac{d g}{d p}
\int_{-1}^{+1} d\mu \frac{1-\mu^2}{\omega+k v(p)\mu \pm \Omega_i}\ .
\label{eq:inte}
\end{equation}

The integral in the variable $\mu$ is now
\begin{equation}
\int_{-1}^{+1} d\mu \frac{1-\mu^2}{\omega+k v(p) \mu \pm \Omega_i}=
{\cal P} \int_{-1}^{+1} d\mu \frac{1-\mu^2}{k v(p)\mu \pm \Omega_i} - i\pi 
\int_{-1}^{+1} d\mu\ (1-\mu^2)\ \delta(k v \mu \pm \Omega_i),
\end{equation}
where $\cal P$ denotes the principal part of the integral and 
we have neglected $\omega$ with respect to $\Omega_i$ (low 
frequency modes). Using Plemelj's formula for the first term one obtains 
for the cosmic ray response:
\begin{eqnarray*}
\chi_{CR} =  \frac{\pi e^2 N_{CR}}{\omega k} \int_{0}^{\infty} dp 
\frac{dg}{dp}
\left[ \left(p^2-p_{min}(k)^2\right) \ln \left\vert \frac{1\pm p/p_{min}}
{1\mp p/p_{min}}
\right\vert
\pm 2 p_{min} p \right] -\\
 - i \frac{\pi^2 e^2 N_{CR}}{\omega k}
\int_{p_{min}(k)}^\infty dp\ \frac{dg}{dp}\ \left(p^2-p_{min}(k)^2\right),
\label{eq:inte1}
\end{eqnarray*}
where we have introduced the minimum momentum $p_{min}(k)=m_i \Omega_i^*/k$, 
which comes from the condition that the second integral in Eq.~\ref{eq:inte} 
is non vanishing only when $\vert\mu\vert\leq 1$, namely when
\begin{equation}
v(p)\geq\frac{\Omega_i}{k}=\frac{\Omega_i^*}{k\gamma}
\Longrightarrow p=\gamma m_i v(p)\geq
m_i \frac{\Omega_i^*}{k}=p_{min}(k).
\end{equation}
The physical
meaning of $p_{min}$ is that of minimum momentum of the protons that can have a
resonant interaction with waves of given wavelength.

In the limit of low frequencies that we are interested in, $\omega+k v_s\ll
\Omega_i^*\ll \vert\Omega_e^*\vert$, the contribution of the background plasma
can be Taylor expanded and the unity in the dispersion relation (displacement
current) neglected. So the dispersion relation reads
\begin{equation}
v_A^2 k^2 = \tilde\omega^2 \pm \frac{N_{CR}}{n_i}(\tilde\omega-k
v_s)\Omega_i^*\left[ 1 \pm I_1^\pm(k) \mp i I_2(k)\right],
\label{eq:disp1}
\end{equation}
where $v_A=B_0/\sqrt{4\pi m_i n_i}$ is the Alfven speed, $\tilde\omega=\omega+k
v_s$ is the wave frequency in the reference frame of the upstream plasma and we
have introduced
\begin{eqnarray}
I_1^\pm (k)&=&\frac{p_{min}(k)}{4} \int_{0}^{\infty} dp \frac{dg}{dp}
\left[ \left(p^2-p_{min}(k)^2\right) \ln \left\vert \frac{1\pm p/ p_{min}}
{1\mp p/p_{min}}\right\vert
\pm 2 p_{min} p \right]\ , \label {eq:Ia1}\\
I_2(k)&=& \frac{\pi}{4} p_{min}(k) \int_{p_{min}(k)}^{\infty} dp
\frac{dg}{dp} \left(p^2-p_{min}(k)^2\right).
\label{eq:Ia2}
\end{eqnarray}
One should notice that the phase velocity of the waves in the plasma
frame is $v_\phi=\tilde\omega/k$ and we want to concentrate on waves
which have a velocity much smaller than the fluid velocity $v_s$
(which is supersonic), therefore $\tilde\omega\ll k v_s$. In this
limit, and using the fact that $I_1^\pm=\pm I_1^+$, one can write the
dispersion relation as 
\begin{equation}
v_A^2 k^2 = \tilde\omega^2 \mp \frac{N_{CR}}{n_i} k v_s \Omega_i^*
\left[ 1 + I_1^+(k) \mp i I_2(k)\right].
\label{eq:disp2}
\end{equation}
This dispersion equation is the same as that found by \cite{bell04} by
treating the background plasma in a MHD approximation. Here we have obtained 
Eq.~\ref{eq:disp2} by assuming that the cosmic ray current and charge are
compensated by a population of cold electrons moving with the cosmic rays, 
a setup which is equivalent to that of having the cosmic ray current of 
\cite{bell04}.
At least within the context of this specific choice of the compensating
current, treating the background plasma within the MHD approximation as
done by \cite{bell04} does not change the results. However the question 
arises of whether the resulting dispersion relation may be different for 
a different and equally reasonable choice of the compensating current. 
In order to investigate this issue, in the 
section below we study the case in which the cosmic ray current is
compensated by assuming a slow drift between thermal ions and electrons. This
approach resembles more closely the kinetic approach first put forward by 
\cite{achter83}. It is also important to stress that this recipe is the same
recently used in the PIC simulations of \cite{niem08}.
 
\subsection{Model B: compensation by electron-proton relative drift motion}
\label{sec:calc2}

The approach described in this section is the one originally put
forward by \cite{achter83}. We show that the dispersion relation is
identical to that found 
in the previous section, provided that the density of cosmic rays is
low enough compared with the density of the gas in the background
plasma. 

Within this approach the electric charge of cosmic rays (assumed to be all
protons) is compensated by the charges of electrons and protons in the
background plasma
\begin{equation}
N_{CR}+n_i=n_e,
\label{eq:charge}
\end{equation}
and the total current induced in the background plasma by the presence of
cosmic rays vanishes, namely
\begin{equation}
0=n_i v_s - n_e v_e.
\label{eq:current}
\end{equation}
This condition can be realized by requiring that electrons and protons
move with slightly different velocities, $v_s$ and $v_e$ respectively. The
small drift between the two species leads to a current which compensates the
cosmic ray current.  

For a given cosmic ray number density $N_{CR}$ the contribution of
accelerated particles to the dispersion relation does not change
compared with the previous model of current compensation. The main
differences with respect to the case presented in the previous section
are that there are no cold electrons, and that electrons and protons
in the background plasma have different velocities and densities. The
contribution of the background plasma to the dispersion relation is
then: 
\begin{equation}
-\frac{4\pi e^2 n_i}{\omega^2 m_i}\frac{\omega+k v_s}{\omega+k v_s\pm
\Omega_i^*} - \frac{4\pi e^2 n_e}{\omega^2 m_e}\frac{\omega+k v_e}{\omega+k
v_e\pm \Omega_e^*}.
\end{equation}
Now, introducing the frequencies $\tilde\omega_i=\omega+k v_s$ and
$\tilde\omega_e=\omega+k v_e$, and using Eqs.~\ref{eq:charge} and
\ref{eq:current}, and again taking the low frequency limit, we obtain
\begin{equation}
-\frac{4\pi e^2}{\omega^2}\left\{
\pm \tilde\omega_i\frac{n_i c}{e B_0} - \left(\frac{\tilde\omega_i}{\Omega_i^*}
\right)^2 \frac{n_i}{m_i} \mp \tilde\omega_e \frac{n_i c}{e B_0}
\mp \tilde\omega_e \frac{N_{CR} c}{e B_0}\right\}.
\end{equation}
Now we notice that 
$$
\pm \tilde\omega_i \mp \tilde\omega_e = \pm k v_s
\frac{N_{CR}}{N_{CR}+n_i}\approx \pm k v_s \frac{N_{CR}}{n_i}
$$
so that, after neglecting terms ${\cal O}\left(\left(N_{CR}/n_i\right)^2\right)$, the
contribution of the background plasma to the dispersion relation becomes:
\begin{equation}
\pm \left(\frac{c}{v_A}\right)^2 \frac{N_{CR}}{n_i}\frac{\Omega_i^*}{\omega}+
\left(\frac{c}{v_A}\right)^2 \left( \frac{\tilde\omega}{\omega}\right)^2.
\end{equation}
It is easy to recognize that the first term is the same as the contribution of
the background plasma in Model A, while the second term is equal to the
contribution of the cold electrons in Model A. It follows that the two
dispersion relations are identical up to linear terms in series
of $(N_{CR}/n_i)$ and start to differ at order ${\cal
  O}\left(\left(N_{CR}/n_i\right)^2\right)$.

At this point it is worth pointing out that in the numerical PIC simulations of
\cite{niem08} the compensating current is realized by assuming a drift between
protons and electrons, exactly as discussed in this section. However, in order
to be able to carry out the calculations, the authors are forced to adopt
unrealistically large values of the ratio $N_{CR}/n_i$ (for the most realistic
cases they use $N_{CR}/n_i=0.3$), which, as discussed above, do
  not necessarily result in a dispersion relation with the same
  characteristics as that found above and derived by \cite{bell04}.

\section{Resonant and non resonant modes}
\label{sec:modes}

In this section we investigate the modes that result from the dispersion
relation in Eq.~\ref{eq:disp2}. For the sake of simplicity we carry out our
calculations for a power law spectrum of accelerated particles with the
canonical shape $g(p)\propto p^{-4}$, which is expected from diffusive
acceleration at strong shocks in the test particle regime. This
regime may potentially be different from the one relevant for
particle acceleration in supernova remnants, in which acceleration
takes place efficiently and the shock becomes cosmic ray modified by
non-linear effects. The most evident effect of this non-linear
reaction is the formation of a shock precursor which reflects in
concave spectra of accelerated particles. The energy content of these
accelerated particles is typically dominated by the highest energy
particles while numerically the low energy particles are still
dominant. On the other hand, efficient acceleration must proceed
parallel to effective magnetic 
field amplification, and as showed by \cite{apjlett,long}, the dynamical
reaction of the strong field leads to a smoothening of the shock
precursor, which in turn leads to spectra of accelerated particles
that though still concave, are closer to power laws. It is straightforward
to grasp the deeply non-linear nature of the problem at hand, where one
is trying to describe the growth of waves associated to cosmic
ray streaming, but the latter only occurs after the waves have 
grown to interesting levels, so as to guarantee efficient particle 
acceleration. At the present time a full analysis of the whole
problem is simply not achievable. In the following we still adopt
the assumption of power law spectra of accelerated particles, and we
point out, where necessary, if a result may be substantially
modified by non-linear effects.

The suitably normalized distribution function $g(p)$ is 
\begin{equation}
g(p) = \frac{1}{p_0^3} \left( \frac{p}{p_0}\right)^{-4}\ \Theta(p-p_0)\ 
\Theta(p_{max}-p)
\label{eq:gp}
\end{equation}
where $\Theta$ is the step function and takes into account the limited
range of momenta spanned by the cosmic ray particles in the
acceleration region. 

The momentum $p_0$ is the minimum momentum of accelerated
particles at the shock location, namely the injection momentum. 
Below, we discuss in detail the dependence of our results on the
choice of the value of $p_0$.

Let us now consider the integrals $I_1$ and $I_2$ in 
Eqs.~\ref{eq:Ia1}-\ref{eq:Ia2}. 
We integrate by parts after performing the substitution 
\begin{equation}
s=\frac{p}{p_{min}(k)}=\frac{p}{p_0}\ \ko,
\end{equation}
obtaining:

\begin{eqnarray}
I_1^+ (k) &=&\frac{1}{4}\frac{p_0^3}{(\ko)^3}
\left\{ 
\left[ g(s) \left((s^2-1) \ln \left\vert \frac{1+ s}{1- s}\right\vert +2 s \right)
\right]_0^\infty-2
\int_0^\infty ds\ s\ g(s)\  \ln \left\vert \frac{1+ s}{1- s}\right\vert
\right\}\\
I_2 (k)&=&\frac{\pi}{4}
\frac{p_0^3}{(\ko)^3}
\left\{
\left[ g(s) (s^2-1) \right]_1^\infty -
2\int_1^\infty ds\ s\ g(s)
\right\}
\label{eq:ints}
\end{eqnarray}

After defining $s_0=\ko$ and $s_2=\ko p_{max}/p_0$ and using the expression 
for $g(s)$ obtained from Eq.~\ref{eq:gp} one finds:
\begin{eqnarray}
I_1^+ (k) &=&-\frac{s_0}{2} \int_{s_0}^{s_2} ds\ s^{-3} 
\ln \left\vert\frac{1+s}{1-s}\right\vert =
-\frac{s_0}{4} \left[\frac{1}{s^2}\left((s^2-1) 
\ln \left\vert\frac{1+s}{1-s}\right\vert-2\ s\right)\right]_{s_0}^{s_2}\\
I_2 (k)&=& \frac{\pi}{2}\ s_0 \left[\frac{s^{-2}}{2}\right]_{s_1}^{s_2}
\end{eqnarray}
 with $s_1={\rm Max}[1,s_0]$.
Finally:

\begin{equation}
I_1^+ (k) =\frac{1}{4 \ko} \left[ \left((\ko)^2-1\right)\ 
\ln \left\vert \frac{1+\ko}{1-\ko}\right \vert -2\ \ko\right]
\label{eq:I1f}
\end{equation}

\begin{equation}
I_2 (k)= -\frac{\pi}{4}
\left\{
\begin{array}{ccc}
\ko &  & \ko \leq 1 \\
(\ko)^{-1} & & \ko \geq 1\ .
\end{array}
\right.
\end{equation}

In terms of the latter, the imaginary and real parts of the frequency can
be written as:
\begin{equation}
\tilde\omega_I^2(k) = \frac{1}{2}\left[-\left(k^2 v_A^2 \pm 
\alpha(1+I_1(k))\right)+\sqrt{\left(k^2 v_A^2 \pm 
\alpha(1+I_1(k))\right)^2 + \alpha^2 I_2^2}\,\,  \right]
\label{eq:omegaI}
\end{equation}
\begin{equation}
\tilde\omega_R(k)=-\frac{\alpha I_2}{2 \tilde\omega_I},
\label{eq:omegaR}
\end{equation}
where $\alpha=\frac{N_{CR}}{n_i}k v_s \Omega_i^*$. 
It is useful to express $\alpha$ as a function of the acceleration efficiency of
the shock. The total pressure in the form of accelerated particles is 
\begin{equation}
P_c = \frac{1}{3}N_{CR}\int_{p_0}^{p_{max}} dp p^3 v(p) g(p) \approx
\frac{1}{3} N_{CR} c p_0 \ln\left(\frac{p_{max}}{m_i c} \right).
\end{equation}
The second part of this expression is a consequence of the fact that
for spectra harder than $p^{-5}$ and not harder than $p^{-4}$, which are
of interest here, the pressure is mainly contributed by relativistic
particles with $p\sim m_i c$. In case of cosmic ray modified shocks, spectra
can becomes mildly harder than $p^{-4}$, but this is not expected to
affect our conclusion in a dramatic way, since it only introduces a
weak dependence on $p_{max}$. Moreover, in case of strongly
modified shocks, the major complication does not come from the
spectrum of accelerated particles, but rather from the fact that the
upstream plasma develops a precursor, namely a gradient in the fluid
velocity that makes the standard treatment illustrated here formally
not applicable. On the other hand, as pointed out several times
throughout the paper, the magnetic reaction of the shock acts in the
direction of smoothening the precursor, so that although formally the
mathematical treatment illustrated here is not applicable, in practice
it should provide a good description of the physical processes at
work.

If we define $\eta=P_c/(n_i m_i v_s^2)$ as the acceleration
efficiency, we can write: 
\begin{equation}
\alpha = 3\eta\frac{1}{R}\frac{v_s^3}{c}\frac{k}{r_{L,0}} = \sigma
\frac{k}{r_{L,0}},
\end{equation}
where $R=\ln\left(\frac{p_{max}}{m_i c} \right)$ and $r_{L,0}=p_0 c/e B_0$ is
the Larmor radius of the particles with momentum $p_0$ in the background
magnetic field $B_0$. We have also introduced
$\sigma=3 \eta\frac{1}{R}\frac{v_s^3}{c}$. A resonant mode can be obtained
from Eq.~\ref{eq:omegaI} with both signs of the polarization. On the other hand
the non-resonant mode only appears when the lower sign is chosen. 

We notice that the following relation holds:
\begin{equation}
\frac{\sigma}{v_A^2}=\frac{N_{CR}}{n_i}\frac{p_0}{m_i c}\frac{v_S
  c}{v_A^2}= \frac{4\pi}{c} J \frac{r_{L,0}}{B_0},
\end{equation}
where $J=e N_{CR} v_S$. This means that the system is strongly current
driven when $\frac{\sigma}{v_A^2}\gg 1$.

\begin{figure}
\resizebox{\hsize}{!}{
\includegraphics{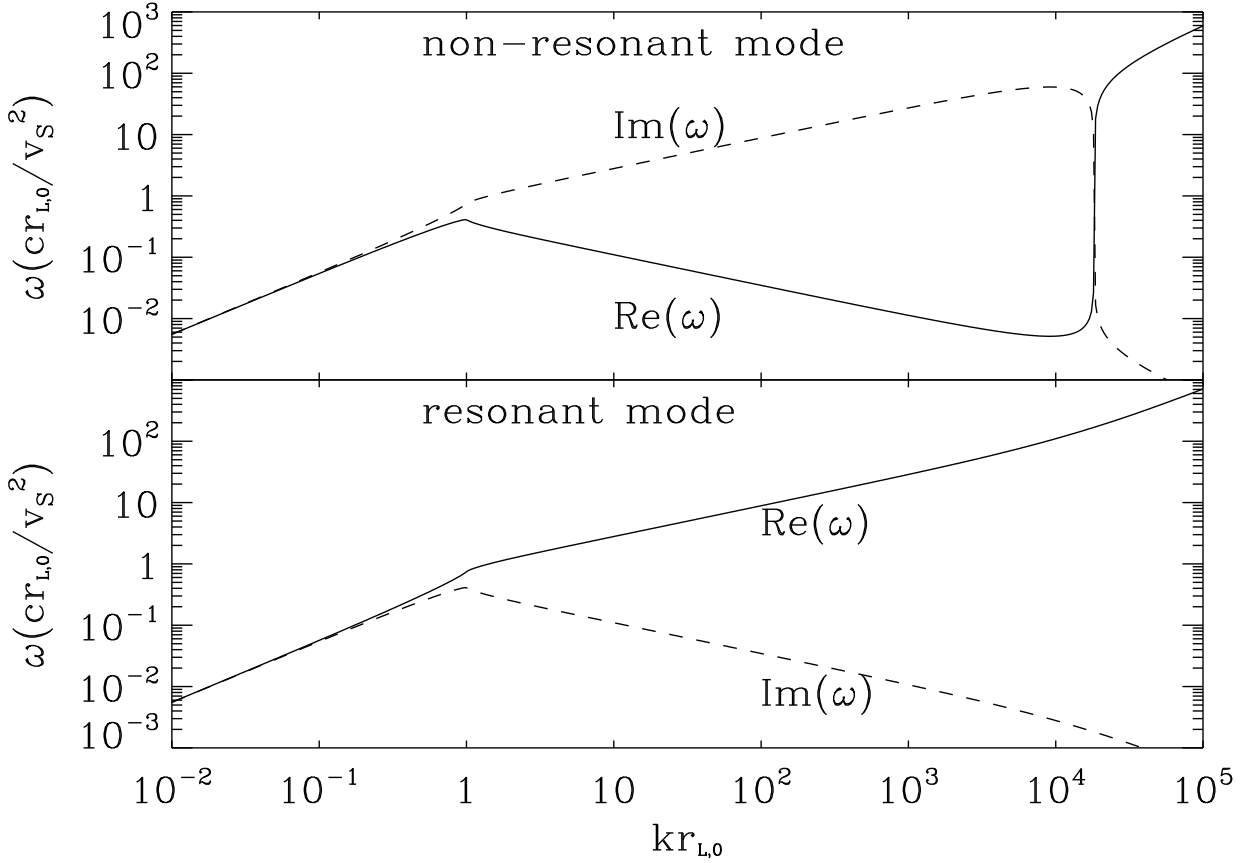}
}
\caption{We plot the real and imaginary part of the frequency as a function 
of wavenumber for the resonant and non-resonant modes. Wavenumbers are in 
units of $1/r_{L,0}$, while frequencies are in units of $v_S^2/(cr_{L,0})$. The
top panel refers to the non-resonant branch, while the lower panel is for the
resonant branch. In each panel, the solid (dashed) curve represents the real 
(imaginary) part of the frequency. The values of the parameters are 
as follows: $v_S=10^9 \rm cm\ s^{-1}$, $B_0=1 \mu G$, $n_i=1\ \rm cm^{-3}$, 
$\eta=0.1$,$p_{max}=10^5\ m_p c$.}
\label{fig:disp9}
\end{figure}

Therefore the parameter $\sigma/v_A^2$ controls the growth rate of the
non-resonant mode: when $\sigma/v_A^2\gg 1$ the non-resonant mode is
almost purely growing and its growth is very fast. When
$\sigma/v_A^2\ll 1$, the non-resonant mode is subdominant and a
resonant mode is obtained, asymptotically identical to that
corresponding to left-hand polarized waves (upper sign in
Eq.~\ref{eq:omegaI}). 

In the following we often refer to the mode
arising with the lower sign of the polarization as the {\it non-resonant
mode}, although one should keep in mind that its peak growth rate reduces to
that of the standard resonant mode in the limit $\sigma/v_A^2\ll 1$. 

In Fig.~\ref{fig:disp9} we plot the solution of the dispersion relation 
in a case for which $\sigma/v_A^2 \gg 1$. The values of the parameters are: 
$v_S=10^9 \rm cm\ s^{-1}$, $B_0=1 \mu G$, $n_i=1\ \rm cm^{-3}$,
$\eta=0.1$, $p_{max}=10^5\ m_p c$.  
The frequency (y-axis) has been normalizated to the advection time for
a fluid element upstream of the shock through the characteristic
distance $cr_{L,0}/v_S$, namely $cr_{L,0}/v_S^2$ . It is worth
stressing however that this is a good estimate of the diffusion time
scale only in the case of Bohm diffusion, when the diffusion
coefficient is $D(p)\approx r_{L,0}c$ and the diffusion time scale is 
$D(p)/v_S\approx cr_{L,0}/v_S^2$.
The plots in the upper (lower) panel in Fig.~\ref{fig:disp9} are
obtained by choosing the lower (upper) sign of the polarization in the
dispersion relation (Eq.~\ref{eq:omegaI} and Eq.~\ref{eq:omegaR}). 

First, let us comment on the consistency of our derivation of the
disperion relation. It is easy to check, from Fig.~\ref{fig:disp9}, that 
these are indeed low frequency modes. More specifically, they satisfy both 
assumptions underlying our calculation: $\tilde \omega \ll k v_S$
and $\tilde \omega \ll \Omega_i$. Moreover, the non-resonant mode (lower
sign of the polarization in the dispersion relation) is 
characterized by an imaginary part that is much larger than its 
oscillatory part for a very large range of wavenumbers. In this same range
of $k$, for our choice of the parameters, its growth is much faster
than for the resonant branch.

Further insight in the behaviour of the different wave modes can be gained
by investigating the limits of the dispersion relation for the
regimes $\ko\ll 1$ and $\ko\gg 1$. Based on these asymptotic trends it
is easy to explain (see \S \ref{sec:p0}) why the normalized
frequencies as functions of the normalized wavenumber, $kr_{L,0}$, do
not depend on the choice of the minimum momentum.

\subsection{Large wavenumber limit: $\ko\gg 1$}
\label{sec:largek}
For $\ko\gg 1$ one easily obtains that $I_1(k)\simeq -(1/3)/(kr_{L,0})^2$ 
and $I_2(k)\simeq -(\pi/4)/ \ko$. In 
Eq.~\ref{eq:omegaI} there are three terms that determine the actual 
dependence of $\tilde\omega_I$ on wavenumber $k$, namely 
$$k^2 v_A^2\propto k^2,\,\,\,\,\,\,\alpha(1+I_1)\approx \sigma
\frac{k}{r_{L,0}}\propto k,\,\,\,\,\,\,\alpha I_2 \approx
-\frac{\pi}{4}\frac{\sigma}{r_{L,0}^2}\propto k^0.$$
In the range $\ko\gg 1$ the third term is always subdominant. Moreover we
can identify two critical values of the wavenumber $k$, $k_1$ and $k_2$, such
that for $k \gg k_1 \gg 1/r_{L,0}$ the first term dominates upon
the third, and for $k \gg k_2$ the first term also dominates upon the 
second term, which is linear in $k$. It is easy to find that 
\begin{equation}
k_1 r_{L,0} = \frac{1}{2} \sqrt{\frac{\pi\sigma}{v_A^2}}
\approx 120
\left(\frac{\eta}{0.1}\right)^{1/2}
\left(\ln\left(\frac{p_{max}}{10^5 m_p c}\right)\right)^{-1/2}
\left(\frac{v_s}{10^9 \rm cm/s}\right)^{3/2}
\left(\frac{B_0}{1 \mu G}\right)^{-1} \left(\frac{n_i}{\rm cm^{-3}}\right)^{1/2}.
\end{equation}
Similarly
\begin{equation}
k_2 r_{L,0} = \frac{\sigma}{v_A^2}\approx
1.8 \times 10^4
\left(\frac{\eta}{0.1}\right)
\left(\ln\left(\frac{p_{max}}{10^5 m_p c}\right)\right)^{-1}
\left(\frac{v_s}{10^9 \rm cm/s}\right)^{3}
\left(\frac{B_0}{1 \mu G}\right)^{-2} \left(\frac{n_i}{\rm cm^{-3}}\right).
\label{eq:k2}
\end{equation}
For $k\ll k_2$ Eq.~\ref{eq:omegaI} gives
\begin{equation}
\tilde\omega_I^2 = \frac{1}{2} \sigma\frac{k}{r_{L,0}} 
\left\{
\mp 1 + 1 + \frac{\pi^2}{32} \frac{1}{(\ko)^2}
\right\}.
\label{eq:kllk2}
\end{equation}
When the upper sign is chosen in the above equation, we obtain
$\tilde\omega_I\propto k^{-1/2}$, while when the lower sign of the polarization
is considered one finds $\tilde\omega_I\propto k^{1/2}$, namely the growth rate 
of the waves increases with $k$. This is the non resonant 
branch found by \cite{bell04}. For this mode $\tilde\omega_I$ increases with $k$
up to $k\sim k_2$ and in the range of wavenumbers between $k_1$ and $k_2$ is
larger than for the resonant waves. The maximum growth rate is obtained for
$k\sim k_2$. In fact, for $k\gg k_2$ one finds:
\begin{equation}
\tilde\omega_I^2 \approx \frac{\pi^2}{64}\ \frac{\sigma^2}{v_a^2 r_{L,0}^2}\ \frac{1}{(k
r_{L0})^2},
\label{eq:kggk2}
\end{equation}
which implies $\tilde\omega_I\propto k^{-1}$ for both the resonant and 
non-resonant modes.

The non resonant mode disappears when $k_1$ becomes larger than $k_2$, which
happens for 
\begin{equation}
\frac{\sigma}{v_A^2}<\frac{\pi}{4} \to \eta < 
4.3 \times 10^{-6}
\left(\ln\left(\frac{p_{max}}{10^5 m_p c}\right)\right)
\left(\frac{v_s}{10^9 \rm cm/s}\right)^{-3}
\left(\frac{B_0}{1 \mu G}\right)^{2}
\left(\frac{n_i}{\rm cm^{-3}}\right)^{-1}.
\label{eq:eta}
\end{equation}

For the reference values of the parameters, the non-resonant mode grows faster
than the resonant mode only for unreasonably low efficiencies of particle
acceleration, as one may conclude by comparing Eq.~\ref{eq:eta} with $\eta\sim
0.1-0.2$ required for the association of cosmic rays to supernova remnants. 

On the other hand, for shock velocity $v_s=10^8 \rm cm\ s^{-1}$ and magnetic
field $B_0=5\mu G$ one easily sees that the limit in Eq.~\ref{eq:eta} becomes
$\eta<0.1$. This implies that the resonant and non resonant modes compete
during the history of a supernova remnant, with the resonant mode prevailing
during the stages in which the shock has slowed down appreciably. We will
comment further on this point below.

We summarize the results of this section by giving the following useful
approximations to the solution of the dispersion relation in the large
wavenumber limit.
For $1/r_{L,0} \ll k_1 \ll k \ll k_2$ we have:
\begin{eqnarray}
\tilde \omega_I^{\rm res} \approx 
\tilde \omega_R^{\rm non-res} \approx 
\frac{\pi}{8}\ \sqrt{\frac{\sigma}{r_{L,o}^3}}\ k^{-1/2} & {\rm and} &
 \tilde \omega_I^{\rm non-res}\approx 
\tilde \omega_R^{\rm res} \approx 
\sqrt{\frac{\sigma}{r_{L,o}}}\ k^{1/2} 
\label{eq:omapp2}
\end{eqnarray}

For $k_2 \ll k$ we have:
\begin{eqnarray}
\tilde \omega_I^{\rm res} \approx
\tilde \omega_I^{\rm non-res} \approx
\frac{\pi}{8}\ \frac{\sigma}{v_A r_{L,0}^2}\ k^{-1} & {\rm and} &
\tilde \omega_R^{\rm res} \approx
\tilde \omega_R^{\rm non-res} \approx k v_A
\label{eq:omapp3}
\end{eqnarray}

\subsection{Small wavenumber limit: $\ko\ll 1$}
\label{sec:smallk}

In the limit of perturbations with wavelength much larger than the gyroradius
of the lowest energy particles in the cosmic ray spectrum, the
results depend again on the ratio between $v_A^2$ and $\sigma$. As we already 
mentioned, for most regions of the parameters space $v_A^2\ll \sigma$, to
which case we limit our analysis here, while we defer to next section a 
discussion of what happens for slow shocks.

For $\ko \ll 1$ one has $I_1(k)\to -1$. Hence, in the limit $v_A^2\ll \sigma$, 
we find for both signs of polarization:
\begin{equation}
\tilde \omega_I\approx\tilde \omega_R\approx \sqrt{\frac{\pi \sigma}{8}}\ k\ .
\end{equation}

\subsection{Dependence on the minimum momentum}
\label{sec:p0}

The number density of accelerated particles at the shock depends on
the minimum momentum $p_0$, which in practice it is hard to know or to
predict, since it is determined by details of the microphysics of the
shock formation. It is therefore useful to address the issue of the
dependence of our results from the assumed value of $p_0$. 

Let us start from the non resonant mode. For $k r_{L,0}\gg 1$,
following Eq. \ref{eq:kllk2}, we have that $\tilde\omega_I^2\propto
k/r_{L,0}$ for $k\ll k_2$ and $\tilde\omega_I^2\propto 1/(k^2
r_{L,0}^4)$ for $k\gg k_2$. It follows that apparently the dependence
on $p_0$ is severe. However, if the wavenumbers $k$ are normalized to
$1/r_{L,0}$ and $\tilde\omega_I$ is normalized to the diffusion time
scale $r_{L,0} c/v_S^2$, the resulting imaginary and real parts of the
frequency are independent
of the choice of $p_0$. In other words, the absolute value of the
growth rate is affected by the choice of $p_0$, but the rate itself,
which is regulated by the time available for the instability to grow,
as measured by the time scale $r_{L,0} c/v_S^2$, is independent of the
choice of $p_0$. Similar considerations can be easily repeated for the
resonant mode and for the low frequency regime of the non resonant
mode. 

\section{Resonant and non-resonant modes in SNRs}
\label{sec:res}
We now study the relative importance of the resonant and non-resonant 
wave modes during the evolution of a SNR. We consider a remnant 
originating in a SN explosion with energy $E_{SN}$. Once the remnant has 
entered the Sedov phase, the shock velocity as a function of time $t$
can be written as:
\begin{equation}
v_S \approx 4 \times 10^8 \rm cm\ s^{-1}\
\left( \frac{E_{SN}}{10^{51} \rm erg}\right)^{1/5}
\left( \frac{n_i}{1 \rm cm^{-3}}\right)^{-1/5}
\left(\frac{t}{10^3 \rm yr}\right)^{-3/5}\ .
\label{eq:sedov}
\end{equation}
As discussed in the previous section, the existence of the non-resonant mode 
depends on the ratio $\sigma/v_A^2$, which can be written as a function of 
the age of the remnant as:
\begin{equation}
\frac{4}{\pi}\ \frac{\sigma}{v_A^2}=1.5 \times 10^3
\left(\frac{\eta}{0.1}\right)
\left(\frac{E_{SN}}{10^{51} \rm erg}\right)^{3/5}
\left(\frac{n_i}{\rm cm^{-3}}\right)^{2/5}
\left(\frac{B_0}{1 \mu G}\right)^{-2}
\left(\ln\left(\frac{p_{max}}{10^5 m_pc}\right)\right)^{-1}
\left(\frac{t}{10^3 \rm yr}\right)^{-9/5}\ .
\label{eq:k1k2}
\end{equation}

\begin{figure}
\resizebox{\hsize}{!}{
\includegraphics{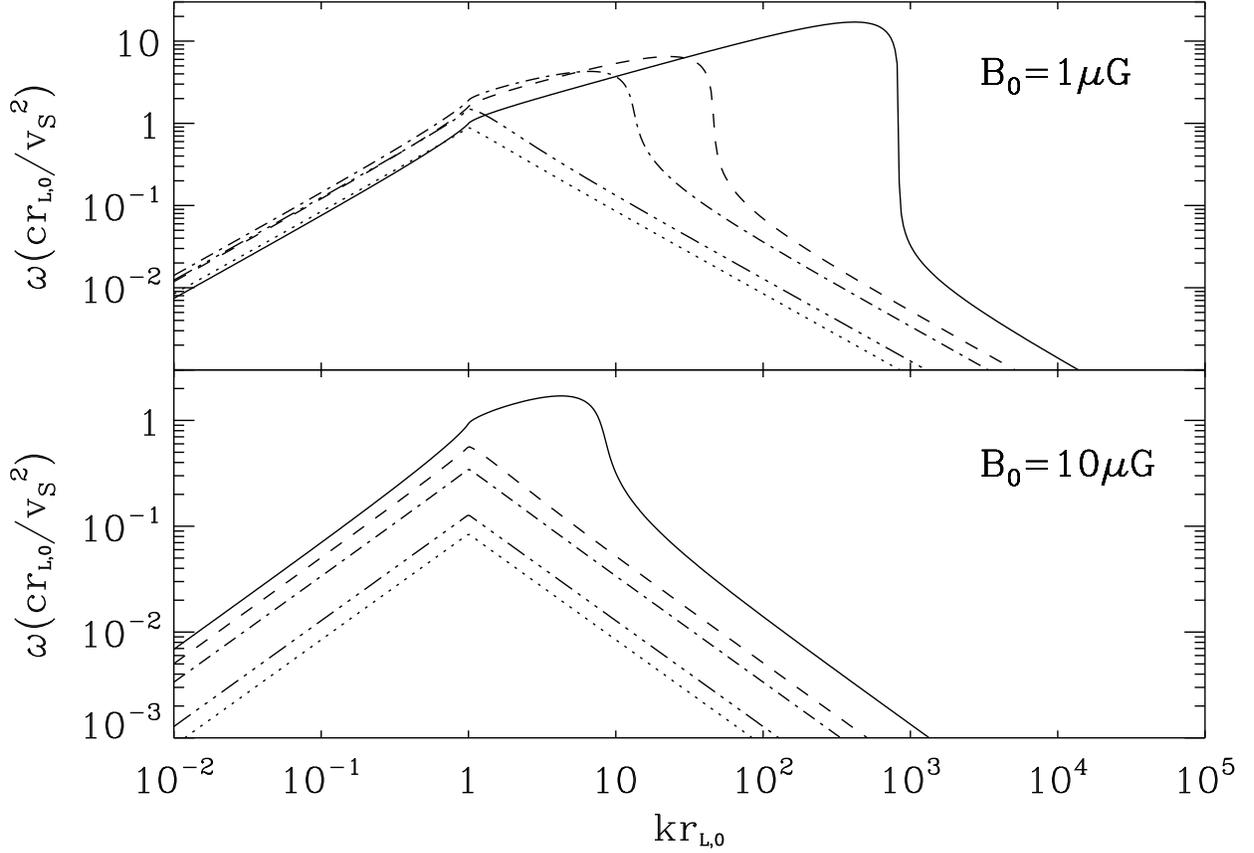}
}
\caption{We plot the growth rate of the non-resonant mode as a function of 
wavenumber. Wavenumbers are in units of $1/r_{L,0}$, while growth-rates are 
in units of the advection time $v_S^2/cr_{L,0}$. The different curves in 
each panel refer to 
different ages of the remnant: solid is for $10^3$ yr, dashed is for 
$5 \times 10^3$, dot-dashed is for $10^4$, dot-dot-dashed is for 
$5 \times 10^4$ and finally dotted is for $10^6$. Two different values
of the background magnetic field strangth are assumed in the two panels:
$B_0=1\ \mu G$ in the top panel, and $B_0=10 \mu G$ in the bottom one.
The shock velocity is computed according to the Sedov expansion of a 
remnant with $E_{SN}=10^{51}$ erg. The remaining parameters are as 
follows: $n_i=1\ \rm cm^{-3}$, $\eta=0.1$, $p_{max}=10^5\ m_p c$.}
\label{fig:evol}
\end{figure}

This implies that during the evolution of a ``typical'' supernova remnant,
Bell's instability, which requires $\sigma/v_A^2 \ge 1$, is likely to 
operate only at early times after the beginning of the Sedov phase.
The non-resonant mode disappears when the remnant is a few 
$10^4$ yr old if it is expanding in a $1 \mu G$ magnetic field
and 10 times faster (age about a few $10^3$ yr) if the background
magnetic field is 10 times higher. At later times, the streaming cosmic 
rays will still amplify the field but only via the classical resonant 
mechanism. This is also clear from Fig.~\ref{fig:evol} where we plot 
the growth rate of the non-resonant mode as a function of age for the 
above mentioned values of the magnetic field: $B_0=1 \mu G$ in the upper 
panel and $B_0=10 \mu G$ in the lower panel. 
From the plots in Fig.~\ref{fig:evol}, where again the time-scale for
wave growth is normalized to the fastest of the time-scales involved 
in the system dynamics, $c r_{L,0}/v_S^2$, one immediately sees that at least
the resonant mode of the streaming instability still grows efficiently after 
$10^6$ yr since the supernova explosion. The non-resonant mode, on the other
hand, soon becomes subdominant. 

\begin{figure}
\resizebox{\hsize}{!}{
\includegraphics{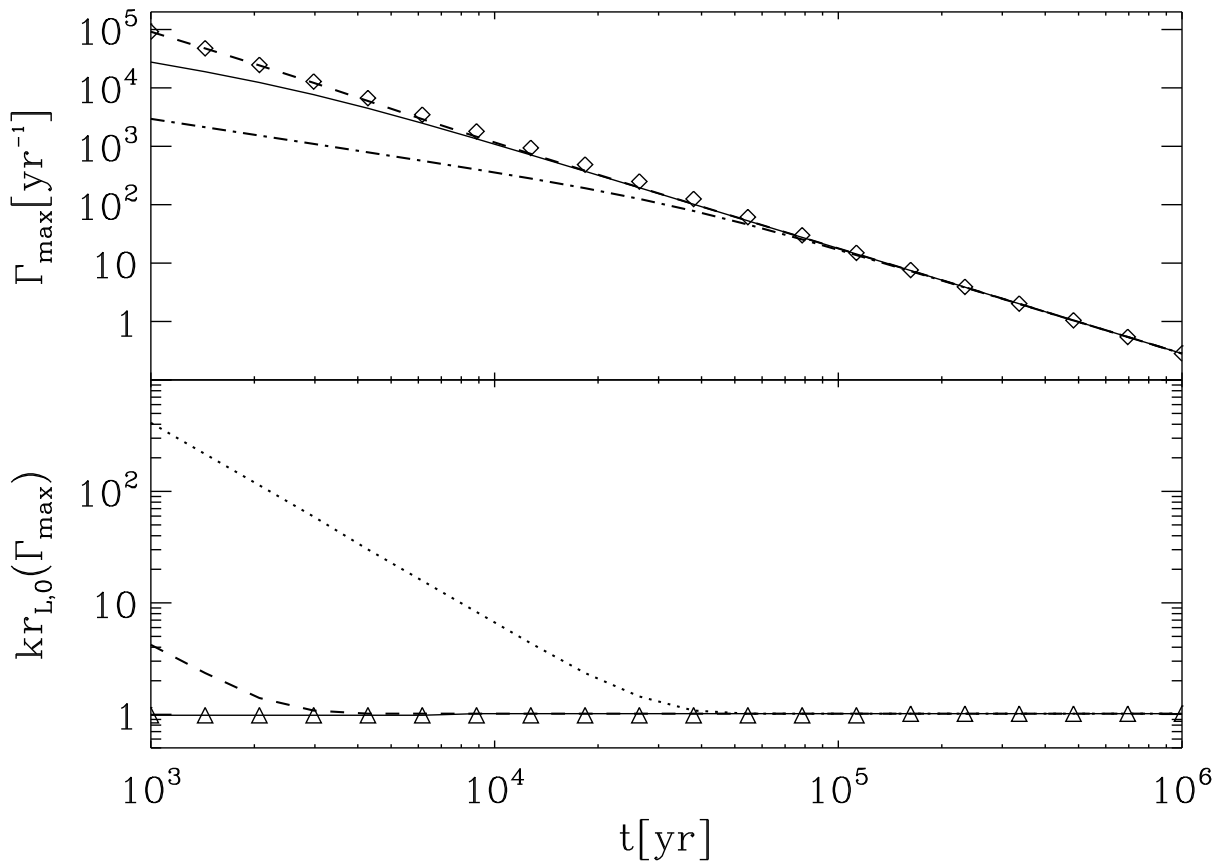}
}
\caption{In the top panel we plot the maximum growth rate of the resonant 
and non-resonant branches as a function of the age of the supernova remnant. 
The growth rate $\Gamma_{\rm max}=\max(\tilde \omega_I)$ is in units of 
${\rm yr}^{-1}$ while along the x-coordinate time is expressed in yr. 
The notation for the
different curves is as follows: the dashed line and the symbols are for the 
non-resonant mode in a $10 \mu G$ and $1 \mu G$ magnetic field respectively;
the solid and dot-dashed lines are for the growth of the resonant mode, 
again for $B_0 = 10 \mu G$ and $B_0=1 \mu G$ respectively.  
In the bottom panel we plot the wavenumber corresponding to the fastest 
growing wave mode for the same situations considered above. Wavenumbers are 
in units of $r_{L,0}$ and the notation for the different line-types is as 
follows: the dashed and dotted lines are for the non-resonant mode in a 
$10 \mu G$ and $1 \mu G$ magnetic field respectively; the solid line and 
symbols are for the resonant mode, again for $B_0 = 10 \mu G$ and 
$B_0=1 \mu G$ respectively. The shock velocity changes with time according 
to the Sedov evolution of a remnant with $E_{SN}=10^{51}$ erg. The remaining 
parameters are as follows: $n_i=1\ \rm cm^{-3}$, $\eta=0.1$, $p_{max}=10^5\ m_p
c$, $p_{inj}=10^{-2}m_i c$.} \label{fig:maxgrowth}
\end{figure}

The non-resonant mode grows the fastest at $k \sim k_2/2$, so that from
Eq.~\ref{eq:omapp2} we can derive the maximum growth rate as:
\begin{equation}
\Gamma_{\rm max}\approx max(\tilde
\omega_I)=\left(\frac{\sigma}{r_{L,0}}\right)^{1/2}
\left(\frac{k_2}{2}\right)^{1/2}=  
\frac{\sigma}{2^{1/2} v_A r_{L,0}}.
\label{eq:maxg}
\end{equation} 
It is clear that, remarkably, $\Gamma_{\rm max}$ does not depend on the
background magnetic field $B_0$. On the other hand, the wavenumber at which
the growth is maximum does depend on $B_0$ (see Eq.~\ref{eq:k2}).
These trends are clearly seen from Fig.~\ref{fig:maxgrowth}, where we
plot the dependence on time of the maximum growth rate, $\Gamma_{\rm max}$, 
and of the wavenumber for which this occurs, for both the resonant and 
non-resonant modes. The plot refers to the ``typical'' SNR parameters 
considered above and the two mentioned values of the background magnetic 
field strength. In both this figure and Fig.~\ref{fig:evol} a 10 \% particle 
acceleration efficiency was assumed, and kept constant during the evolution of
the remnant. This latter assumption is definitely not very realistic, as can be
demonstrated by using non-linear theory of particle acceleration.

In Fig.~\ref{fig:maxgrowth} the time at which the fastest growing mode switches
from non-resonant to resonant is identified by the intersection between the
dashed line and the solid ($B_0=10 \mu G$) or the dot-dashed ($B_0=1 \mu G$) 
one depending on the magnetic field strength. The dominant wave mode 
progressively moves to larger wavelengths. The implications of this peculiar
trend are expected to be profound on the determination of the diffusion
coefficient: we recall that the standard Bohm diffusion is the limit obtained
for resonant interactions of particles and waves when $\delta B(k)=B_0$ for any
value of $k$. For non-resonant modes, the diffusion properties need to be
recalculated from first principles. On one hand, since the most unstable modes
have $k\gg 1/r_{L,0}$, most particles do not resonate with these modes and the
typical deflection suffered by a single particle within a spatial scale $\sim
1/k$ is very small. On the other hand the number of scattering events is very
large, therefore a substantial reduction of the diffusion coefficient can still
be expected (see \cite{reville08} and \cite{zira08b}).

\section{Conclusions}
\label{sec:concl}

We investigated the excitation of streaming instability induced by accelerated
particles in the vicinity of a non-relativistic shock wave, typical of
supernova shells expanding in the interstellar medium. The calculation is based
on kinetic theory, hence we do not require the MHD approximation to hold for
the background plasma. We find that the dispersion relation of the
waves leads to the appearance of two modes, a resonant and a non-resonant one.
The former is the well known unstable mode, discussed by
\cite{zweib79} and \cite{achter83}, based on a resonant interaction between waves
and particles. The latter is similar to that discussed by \cite{bell04}, who
however based his analysis on a set of assumptions that called for further
investigation: the calculation of \cite{bell04} is based on the assumption that 
the background plasma can be treated in the MHD regime, and makes specific
prescriptions on the return current which compensates the cosmic ray current
upstream of the shock. Moreover, the whole calculation is carried out in the 
frame of the upstream plasma, where in principle there is no stationary solution 
of the problem.

Our kinetic calculations are carried out for two models of the compensating
current: in the first model, the return current is established through a
population of cold electrons, at rest in the shock frame, which exactly
compensate the positive charge of cosmic ray protons. In the second model, the
return current is due to a slight drift between ions and electrons in the
background plasma upstream of the shock. We have demonstrated that the dispersion
relation of the waves is the same in the two cases, to order
${\cal O}\left(N_{CR}/n_i\right)^2$. 

The resonant and the non-resonant mode are found at the same time, with growth
rates which in the general case are different. The non-resonant mode is almost
purely growing and is very apparent when particle acceleration is efficient.
The parameter that regulates the appearance of the non-resonant mode is
$\sigma/v_A^2$, where $\sigma=3\eta v_S^3/(c R)$. When $\sigma/v_A^2\gg 1$, the
waves excited in a non-resonant way grow faster than the resonant modes and may
lead to a substantial magnetic field amplification. 

The strong dependence of $\sigma$ on the shock velocity implies that the
non-resonant mode is likely to be the dominant channel of magnetic field
amplification in SNRs in the free expansion phase and at the early stages of
the Sedov-Taylor phase of adiabatic expansion. At later times, the non-resonant
mode {\it collapses} on the resonant mode, which keeps providing appreciable
growth for longer times, at least if damping mechanisms are neglected.
The growth of the fastest non-resonant mode is independent on the strength of
the unperturbed initial magnetic field $B_0$.

The non-resonant mode, when present, grows the fastest at wavenumber $k_2$
given by Eq. \ref{eq:k2}, which in the cases of interest is much larger than
$1/r_{L,0}$, where $r_{L,0}$ is the gyroradius of the particles with minimum
momentum in the cosmic ray spectrum. These modes are therefore
short wavelength waves, which is the main reason why the assumption of
stationarity in the upstream frame, as required by \cite{bell04}, was
acceptable, despite the impossibility of reaching actual stationarity in that
frame. 

The numerical results in this paper were specialized to the case of a power law
spectrum $p^{-4}$ of accelerated particles, typical of Fermi acceleration
at strong shocks. However, one should keep in mind that the levels of
efficiency required for the non-resonant mode to appear are such that the
dynamical reaction of the accelerated particles on the shock cannot be
neglected (see \cite{maldrury} for a review). This backreaction leads to several
important effects: on one hand the spectra of accelerated particles become
concave, and concentrate the bulk of the energy in the form of accelerated
particles at the maximum momentum. On the other hand, the efficiently amplified
magnetic field also exerts a strong dynamical reaction on the system, provided
the magnetic pressure exceeds the gas pressure in the shock region
(\cite{apjlett,long}). 
This second effect results in an enhanced acceleration efficiency (due
to large B-fields) but weaker shock modification (spectra closer to power laws)
due to the reduced compressibility of the plasma in the presence of the 
amplified magnetic field. These non-linear effects cannot be taken
into account in the type of calculations presented here, although we
do not expect the qualitative character of our conclusions to be
affected profoundly by them.

Nevertheless it is useful to go through the possible consequences of
the non-linear effects in a somewhat deeper detail: one can expect two
types of complications, one of principle and the other in the
numerical values of the growth rates. The latter simply derives from
the approximations intrinsic in the assumptions we made: for instance
the adoption of a power law spectrum of accelerated particles with
slope $4$, which clearly fails when a precursor is formed.

There are then complications coming from deeper unknown pieces of
Physics or from a not totally satisfactory mathematical approach. For
instance, the standard perturbative approach adopted here is based on
the assumption of a spatially uniform background. The presence of a
precursor invalidates this assumption, although probably not in a
dramatic way.  

Since the non-resonant mode appears for large values of $k$, the
relevant quantities can be assumed to be spatially constant in the
precursor on scales $\sim 1/k$, so that in this respect our
calculations are still expected to hold, and probably to a better
accuracy for the non-resonant modes ($k \gg 1/r_{L,0}$) than for the
resonant ones ($k \sim 1/r_{L,0}$). Moreover, as stressed above, the 
dynamical reaction of the magnetic field leads to weaker modification
of the shock, and therefore to spectra with less prominent concavity  
(closer to $p^{-4}$). Also in this respect, the calculations presented
here should serve as a good description of all relevant physical
effects related to the growth of the cosmic ray induced instabilities.  

More important, the acceleration process is directly affected by the
physics of particles' diffusion in the shock region, which in turn is
determined by the excited waves. This intrinsic non-linearity cannot
be taken into account in perturbative approaches like ours or like
Bell's, and one should always be aware of this limitation. Even more,
while the diffusion coefficient for resonant modes can at least be
derived in quasi-linear theory, at present there is no derivation of
the diffusion coefficient associated with scattering on non-resonant
modes (see \cite{zira08b} for a first attempt at discussing this
effect).

Another issue that deserves further investigation is that of determining
the level of field amplification at which the instability saturates. This
cannot be worked out within a linear theory calculation and only numerical
simulations can address this issue. Recent efforts in this direction have 
been made by \cite{bell04} and \cite{zira08a} through MHD simulations and by
\cite{niem08} by using PIC simulations. While the first two papers find a
saturation level $\delta B^2/(4 \pi) \sim (v_s/c) P_c$, in the third paper a
much lower level of field amplification is found. The authors conclude
that the existence of large magnetic field amplification through the
excitation of non-resonant modes is yet to be established. 

Although we agree with this conclusion, we also think that the setup
of the PIC simulation by \cite{niem08} is hardly applicable to
investigate the excitation of the Bell instability at shocks, or at
least several aspects of it should be 
studied more carefully. First, they carried out the calculations in a
regime in which the condition of strong magnetization,
$\omega\ll\Omega_0$, was violated. Second, in order to carry out the
calculations, 
\cite{niem08} are forced to assume unrealistically large values for
the ratio $N_{CR}/n_i$ (of order 0.3 for their most realistic runs).
The return current as assumed by \cite{niem08} corresponds to our
second model, which however leads to the same dispersion relation as
\cite{bell04} only at order ${\cal O}\left(N_{CR}/n_i\right)^2$, which
is not necessarily the case here. Moreover, the spectrum of accelerated
particles is assumed to be a delta-function at Lorentz factor 2, instead of a
power law (or more generally a broad) spectrum. It is not obvious that 
for non-resonant modes this assumption is reasonable. But the most serious 
limitation of this PIC simulation is in the fact that the authors do not 
provide a continuous replenishment of the cosmic ray current, which is instead 
depleted because of the coupling with waves. In the authors' view this seems 
to be a positive aspect of their calculations, missed by other approaches, 
but in actuality the cosmic ray current is indeed expected to be stationary 
upstream, and we think that the PIC simulation would show this too if 
particles were allowed to be accelerated in the simulation box instead 
of being only advected and excite waves. Clearly if this were done, the 
spectrum of accelerated particles would not keep its delta-function shape, 
but should rather turn into a power-law-like spectrum. The latter issue 
adds to the absence of a replenishment of the current, which seems to us 
to be the main shortcoming of these simulations. Overall, it appears that 
the setup adopted by \cite{niem08} would apply more easily to the
propagation of cosmic rays rather than to particle acceleration in the
vicinity of a shock front. 

The issue of efficient magnetic field amplification, possibly induced by cosmic
rays, has become a subject of very active debate after the recent evidence of
large magnetic fields in several shell-type SNRs. The implications of such
fields for particle acceleration to the knee region, as well as for the
explanation of the multifrequency observations of SNRs, are being investigated.
Probably the main source of uncertainty in addressing these issues is the role
of damping of the excited waves. For resonant modes, ion-neutral damping and
non-linear Landau damping have been studied in some detail: their role depends
on the temperature of the upstream plasma and on the shock velocity. For
non-resonant modes, being at high $k$, other damping channels could be
important (Everett et al., in preparation). Whether SNRs can be the source of
galactic cosmic rays depends in a
complex way on the interplay between magnetic field amplification, damping,
particle scattering and acceleration, together with the evolution of the
remnant itself. 

\section*{Acknowledgments}
The authors are very grateful to J. Everett and E. Zweibel for a
critical reading of the manuscript and for ongoing collaboration and to
B. Reville for constructive discussions. They
also thank J. Kirk for identifying a problem in a previous version of
the manuscript in the low-k part of the dispersion relation. 
This work  was partially supported by PRIN-MIUR 2006, by ASI through
contract ASI-INAF I/088/06/0 and (for PB) by the US DOE and by NASA grant
NAG5-10842. Fermilab is operated by Fermi Research Alliance, LLC under Contract
No. DE-AC02-07CH11359 with the United States DOE.

\bibliographystyle{mn2e}
\bibliography{nldsa}
\end{document}